\documentclass{elsart}
\usepackage{latexsym}
\usepackage{epsfig}
\usepackage{natbib}

\def\url#1{#1} 

\begin{document}

\begin{frontmatter}

\title{The standard theory of extinction and the spectrum of 
stars with very little reddening.}

\author{Fr\'ed\'eric \snm Zagury}
\address{
\cty 02210 Saint R\'emy Blanzy, \cny France
\thanksref{email} }

   \thanks[email]{E-mail: fzagury@wanadoo.fr}

\received{May 2001}
\accepted{July 2000}
\communicated{L.L. Cowie}

 \begin{abstract}
This paper examines the relationship between spectra of stars of 
same spectral type with extremely low reddenings.

According to the standard theory, the relationship between the spectrum of stars 
with same spectral type and small, but different reddenings should be 
different in the optical and in the UV.

This difference is not observed: 
the ratio of the spectra of two stars in directions where the 
reddening is large enough to be detected and low enough not to give a 
noticeable $2200\,\rm \AA$ bump
is an exponential of $1/\lambda$ from the near-infrared to the far-UV.

This result is in conformity with the ideas introduced in preceding 
papers: 
the exponential optical extinction extends to the UV, and the 
spectrum of stars with enough reddening is contaminated by light 
scattered at close angular distance from the stars.

An application will be the determination of the spectrum of a 
non-reddened star from the spectrum of a star of same spectral type with little reddening.
\end{abstract} 
\end{frontmatter}
 \section{Introduction} \label{intro}
The paper compares the spectra of eleven stars of same spectral type 
and very low $E(B-V)$ (a few 0.01).
The UV spectrum of these stars decreases steadily with 
wavelength, with no evidence of the usual 
signs of consequent reddening: there is no apparent $2200\,\rm\AA$ bump 
and no sign of extinction in the far-UV.  
Therefore, these stars are potential candidates to serve as `reference 
stars' to establish the extinction curve in directions where the UV 
extinction features are more obvious.

But the average slope of the spectrum of these stars differs from star 
to star, in the optical and in the UV (section~\ref{analys}).
The variation of slope can either be due to different amounts of 
interstellar matter on the line of sight or to 
variations of stellar properties (temperature for instance).

In the former case the question of how to precisely determine 
the spectrum of a non reddened star of given spectral type arises.
A first step towards the determination of the spectrum of a 
non-reddened star will be to understand the relationship 
which exists between the spectrum of two stars of same spectral type which are reddened but 
not yet enough to present evidence of a bump at $2200\,\rm\AA$.

All observations agree on a linear extinction law in the optical.
The optical spectra of two stars with different reddenings and same 
spectral type will differ by an exponential of $1/\lambda$, 
$e^{2\Delta E/\lambda}$, where $\Delta E$ is the difference in 
$E(B-V)$ ($B-V$ since the stars have same spectral type) 
between the stars and $\lambda$ is in $\mu$m.

\begin{figure}
\resizebox{!}{0.8\columnwidth}{\includegraphics{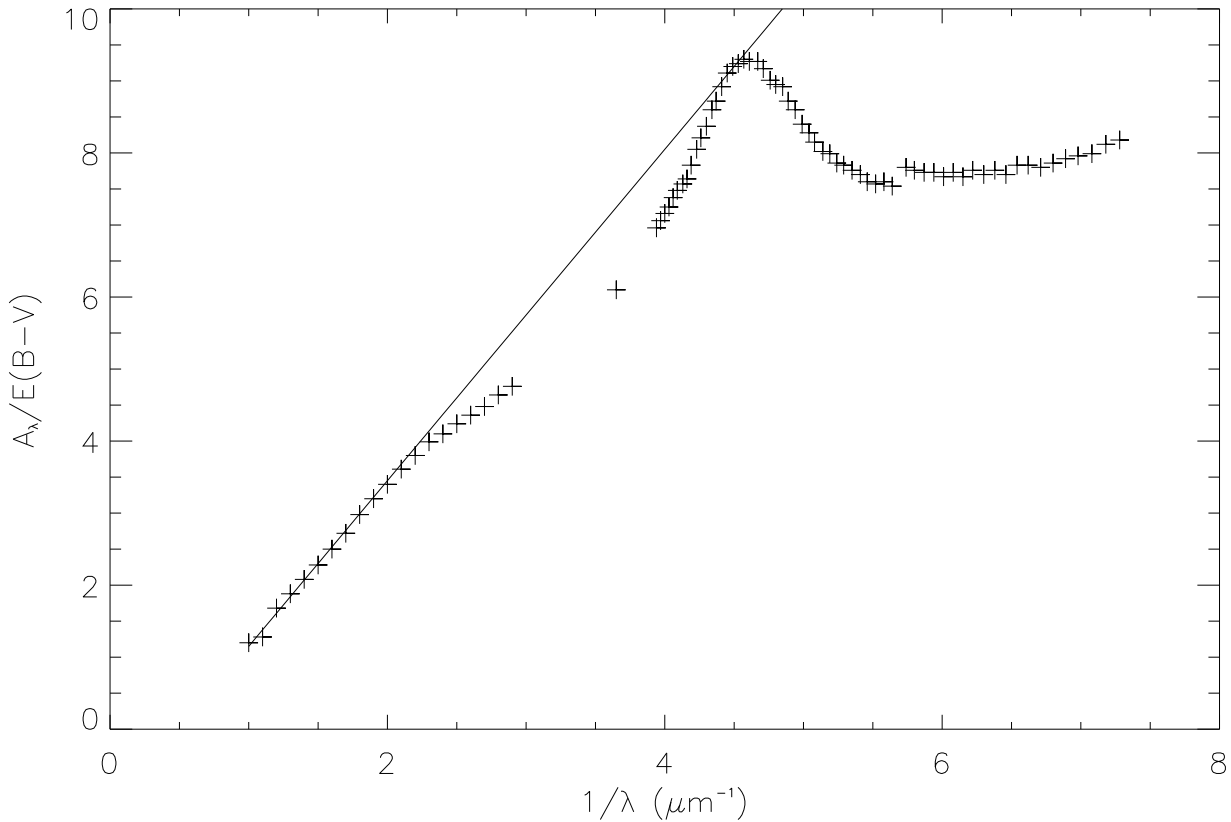}} 
\caption{Seaton's average extinction curve  from  \citet{nandy76}.
The straight line is the linear optical extinction and its' extension 
in the UV.
The two extinction curves are equal in the optical and diverge 
noticeably in the UV.} 
\label{fig:seaton}
\end{figure}

According to the standard theory of extinction, this relationship does not extend to the UV.
Compared to the extension in the UV of the linear optical extinction the 
standard theory predicts an important diminution of extinction in the far-UV.
The change of slope of the extinction from the optical to the UV is 
illustrated in figure~\ref{fig:seaton} where
the average extinction predicted by the standard theory and the 
optical linear extinction and its' extension to the UV are plotted.

On the other hand, I tried to 
call attention  (UV1, UV2, UV3, UV4, see the bibliography) 
to the fact that the observed departure of the UV extinction 
from the straight line drawn on figure~\ref{fig:seaton} can be due to 
the introduction of an additional component of starlight scattered at very small angular 
distance from the star.
This idea was first introduced by \citet{savage75}, and discarded for reasons which are questionable (UV2).
The scattered light is more noticeable in the UV than in the optical
because the extinguished starlight, (therefore the number of photons available 
for scattering) is 
greatly enhanced (as $e^{2E(B-V)/\lambda}$, if the optical extinction 
extends to the UV) at UV wavelengths.

In this case the linear in $1/\lambda$ optical extinction
extends with the same law in the UV, in conformity with 
the basic ideas of the theory of extinction by dust particles.
If extinction is strong enough to be detected, but low enough so 
that the scattered light is negligible, the 
observed extinction law must be close to a continuous exponential decrease from 
the optical to the far-UV.
Within this framework, the ratio of the spectra of two stars with same spectral type and very 
small reddenings is expected to be an exponential of $1/\lambda$ with 
the same exponent over all the optical and UV spectral range.

Therefore the comparison of the spectra of stars with little reddening 
and same spectral type has two outcomes:
\begin{itemize}
    \item  It leads to the determination of the spectrum of a 
    non-reddened star of same spectral type.
    \item  It is an opportunity to compare the predictions of the standard 
    theory to observations in the particular case of regions of very low column densities.
\end{itemize}¥
These are the objectives of the paper.

Section~\ref{data} presents the data used in the paper.
Data are analysed section~\ref{analys}, in view of 
understanding the relation between the spectrum of stars of same 
spectral type and very low reddenings.
In section~\ref{redori} two possible 
causes for the variation of the slope of the spectrum from star to star are reviewed.
From section~\ref{extlaw} on, reddening is assumed to be the main reason of 
these variations.
The extinction law in mediums of very low column density is deduced 
(section~\ref{extlaw}), compared to the standard theory 
(section~\ref{sttheo}), 
and to the alternative explanation of the UV extinction curve 
mentionned above (section~\ref{exptheo}).
Section~\ref{refspec} concludes on a method to obtain unreddened spectra from 
slightly reddened stars.
A summary is given in section~\ref{conc}. 
\section{Data} \label{data}
    \begin{table}[tp]
       \[
    \begin{tabular}{|c|l|c|c|c|c|c|c|}
\hline
$n^\circ$&name &$V \,^{\,(1)}$&$B-V \,^{\,(1)}$&$R-V\, ^{\,(2)}$&
$I-V \,^{\,(1)}$& $2\Delta E_{vis}\,^{\,(3)}$ &$2\Delta E_{uv}\,^{\,(4)}$  \\
\hline
0 & HD31726 &6.15& $-0.202\pm0.004$&-&$0.21\pm 0.01$ &0&0 \\
\hline
1 & HD122980 & 4.36& $-0.198\pm 0.015$& $0.093$ & $0.21\pm 0.02$&
0.006 & 0.006 \\
\hline
2 & HD36166 & 5.77 & $-0.192\pm 0.006$& -& $0.20\pm 0.03$&
0.018 & $0.018$\\
\hline
3 & HD108483 &3.91&$-0.192\pm 0.008$& 0.087& $0.20\pm 0.02$&
0.018 & 0.018 \\
\hline
4 & HD54669 & 6.65 & $-0.189\pm 0.004$ & - & $0.20\pm 0.01$ &
0.025 & $0.030$ \\
\hline
5 & HD192273 & 8.84 & $-0.187\pm 0.012$ & 0.082 & $0.19\pm 0.02$ & 
0.041 &$0.037$\\
\hline
6 & HD42690 & 5.06 & $-0.201\pm 0.007$ & - & $0.21\pm 0.01$ &
0.002 &0.040  \\
\hline
7 & HD64503 & 4.49 & $-0.188\pm 0.008$ & 0.069 & $0.18\pm 0.02$ & 
0.041 &$0.042$\\
\hline
8 & HD110879 & 3.04&$-0.178\pm 0.005$& 0.078 & $0.19\pm0.02$&
0.044 & $0.045$ \\
\hline
9 & HD120086 & 7.87 & $-0.183\pm0.012$ & 0.079 &$0.18 \pm 0.02$ &
0.044 &$0.045$\\
\hline
10 & HD64802 & 5.48 & $-0.175\pm 0.003$ & - & $0.18 \pm 0.00 $ & 
0.053 & $0.053$\\
\hline
11 & HD37397 & 6.84 & $-0.152\pm 0.003$ & - & $0.15\pm 0.00$ & 
0.110 & $0.094$  \\
\hline
\end{tabular}    
    \]
\begin{list}{}{}
\item[$1$] $V$ magnitude, $B-V$ and $I-V$ colors in the Cousins 
system from Hipparcos.
\item[$2$] $R-V$ color in the Cousins system from the Lausanne 
Institut of Astronomy database. 
\item[$3$] $e^{2\Delta E_{vis}}$ times the optical spectrum of the 
star gives the best fit to the spectrum of HD31726.
The error on $2\Delta E_{vis}$  is $\pm 0.005$ on the average.
\item[$4$] $e^{2\Delta E_{uv}}$ times the UV spectrum of the 
star gives the best fit to the UV spectrum of HD31726.
The error on $2\Delta E_{uv}$  is $\pm 0.01$ on the average.
\end{list}
\caption[]{The table is commented in more detail section~\ref{table}.}		
\label{tbl:stars}
\end{table}
\subsection{Star selection} \label{dataselect}
Stars were selected on the basis of three criteria: 
same spectral type, as little reddening as possible,
and to have been observed in the UV.

The spectral type of the stars is B2V.
There is no specific reason for this choice:
it was a search for non-reddened B2V stars that suggested the idea of the 
paper.
The same work can be repeated with other stellar types, with even 
greater facility for O or B0 stars.

The stars were selected from the Hipparcos main catalogue which can 
be queried from Vizier (http://vizir.u-strasbg.fr/) at the Centre de Donnees astronomiques 
de Strasbourg (http://cdsweb.u-strasbg.fr/).
Vizier allows to query the catalogue for stars of a given spectral 
type and to order the stars by increasing $B-V$.
The final selection was done upon whether or not a UV spectrum of 
the star was available.
\subsection{Optical data} \label{dataopt}
The paper relies on the comparison of the average slope of the spectra 
in the different wavelengths ranges (optical and UV).
Since the optical spectra of the 11 selected stars are not available, they have been replaced 
by points on the spectra, given by the photometric observations of 
the stars.
The variation from star to star of the slope of the optical spectrum  
is most easily perceived if the data points are dispatched on a as large as possible wavelength 
range:
the $I$-band on the 
near-infrared side and the $B$-band before the Balmer jump. 

The $B$-band magnitude, the $B-V$ color in the Johnson system and the 
$V-I$ color ($I$ magnitude in the Cousins system) are found in the Hipparcos 
catalogue along with the uncertainty of these values.
For some stars the $R-V$ color (Cousins $R$ magnitude)
is available at the Lausanne Institute 
of Astronomy databases (http://obswww.unige.ch/gcpd/), 
giving a fourth point on the spectra.
The $U$ magnitude, although it can be retrieved for all the stars, was 
not used because it is situated on the Balmer jump, and will introduce 
a random noise in the data.
\subsection{UV data} \label{datauv}
Two sources of UV data were used.
The spectral observations of the International Ultraviolet Explorer (IUE) 
were presented in UV1.
These observations provide high resolution and sensitive spectra,
although problems are often met at the edges of the spectra.
The TD1 mission gives spectra \citep{jamar} with less resolution ($20\,\rm\AA$) but 
with good signal/noise ratio .
TD1 also gives, for a larger number of stars, photometric fluxes on four 
UV bands centered at $\rm 2740\,\AA$, $\rm 2365\,\AA$, $\rm 
1965\,\AA$, $\rm 1565\,\AA$  \citep{thompson}.
The relative error on these fluxes is $\sim 3\, 10^{-3}$ on average.
    
IUE spectra were retrieved as fits files from the IUE website: http://ines.vilspa.esa.es.
All the stars but three have long and small wavelengths range UV
observations.
HD36166 was observed with the LWR camera only, HD37397 with the SWP camera 
only.
HD64503 was observed by IUE with the SWP camera only.
The TD1 spectrum, available for this star, was preferred because of its' 
larger spectral coverage.

It happens that some IUE spectra of the same star and in the same 
wavelength range (long or short) are undercalibrated.
The highest calibration was systematically chosen as the reference one.
For the final UV spectra, continuity between the long and short wavelengths spectra was 
verified after smoothing the spectra with a seven point median filter.
No scaling factor between the long and short wavelength ranges was 
necessary to adjust the spectra for the sample of stars.

The good calibration and the reliability of the spectra was verified 
by a comparison with the TD1 spectra and/or UV photometric bands.
TD1 and IUE agree for all stars except for HD192273 for which IUE 
calibration is lower than TD1 one's by a factor 0.82.
The long wavelength IUE spectra of HD122980 have differences from 
spectrum to spectrum.
I have adopted the one which best matches the slope indicated by the 
TD1 photometry.

The UV spectrum of the stars are plotted (dotted lines referred as (1)), 
scaled by an arbitrary factor and smoothed by a 5 points median filter, 
figures~\ref{fig:uv1} to \ref{fig:uv3}.
\subsection{Table~\ref{tbl:stars}} \label{table}
Table~\ref{tbl:stars} contains the optical data (first 
six columns) and the difference of slope with the spectrum of HD31726 
found for each star in the optical and in the UV (two last columns).

Following are a few comments on the table:

\begin{itemize}
    \item  Column 1: number associated to each star and used to 
    pinpoint the stars on the figures. 
    Stars are classified by decreasing UV slope order 
    (increasing $\Delta E_{uv}$).

    \item  Column 2: name of the star.

    \item  Column 3: $V$ magnitude of the star.
    The $V$ magnitudes are needed for figure~\ref{fig:uvsuri} only 
    (section~\ref{exptheo}). 

    \item  Column 4 to 6: colors of the stars in the Johnson ($B-V$) 
    and  Cousins ($R-V$ and $I-V$) systems.
    The central wavelengths for the $B$, $V$, $R$, $I$ bands were 
    calculated from the filters' respons given at the Lausanne Institut of 
    Astronomy database. 
    The adopted central wavenumbers are:
    $[2.27, 1.8, 1.56, 1.27]\,\mu\rm m^{-1}$.
    The optical photometry is used to derive the best correction 
    ($2\Delta E_{vis}$) to 
    apply to the optical spectrum of one of the stars so that it 
    matches the spectrum of HD31726.
    This correction was checked to be coherent with other photometric 
    systems, uvby and Johnson UBVRI when available.
    uvby photometry, available for all the stars of the sample, is 
    not very useful because of a limited extent in  wavelength.
    UBVRI photometry is available for 4 stars only.
    Wavenumbers also interven in the calculation of the $2\Delta E_{vis}$.
    Possible errors on the central wavelengths were checked to be small enough 
    not to affect the results of the paper.
       
    \item  Column 7: difference of the optical slope between the star and 
    HD31726. Obtained in section~\ref{analyvis} from the colors 
    of the star (figure~\ref{fig:vis}).
    $2\Delta E_{vis}$ is given with an uncertainty of $\pm 0.005$.
    
    \item  Column 8: difference of UV slope between the star and 
    HD31726 obtained in section~\ref{analysuv} through the comparison 
    of the UV spectrum of the stars with the spectrum of HD31726
   (figures~\ref{fig:uv1} to \ref{fig:uv2}).
   The uncertainty is e.g. of an order of $\pm 0.01$.
   The uncertainty for HD36166 and for HD37397 is larger ($\sim 0.020$) because half of the UV
   spectrum only is available. 
   The difference of slope between the spectra of HD36166 and HD122980 
   with the spectrum of HD31726 is extremely small and the three stars 
   can be considered to have nearly equal reddenings.
   
\end{itemize}¥
\section{Data analysis} \label{analys}
All stars will be compared to one star in the optical and in the UV.
The reference or comparison star which is chosed is HD31726 which, 
according to table~\ref{tbl:stars}, has the smallest values of 
$B-V$ and $V-I$ and probably the smallest reddening.
\subsection{UV data} \label{analysuv}
\begin{figure*}[p]
\resizebox{!}{1.5\columnwidth}{\includegraphics{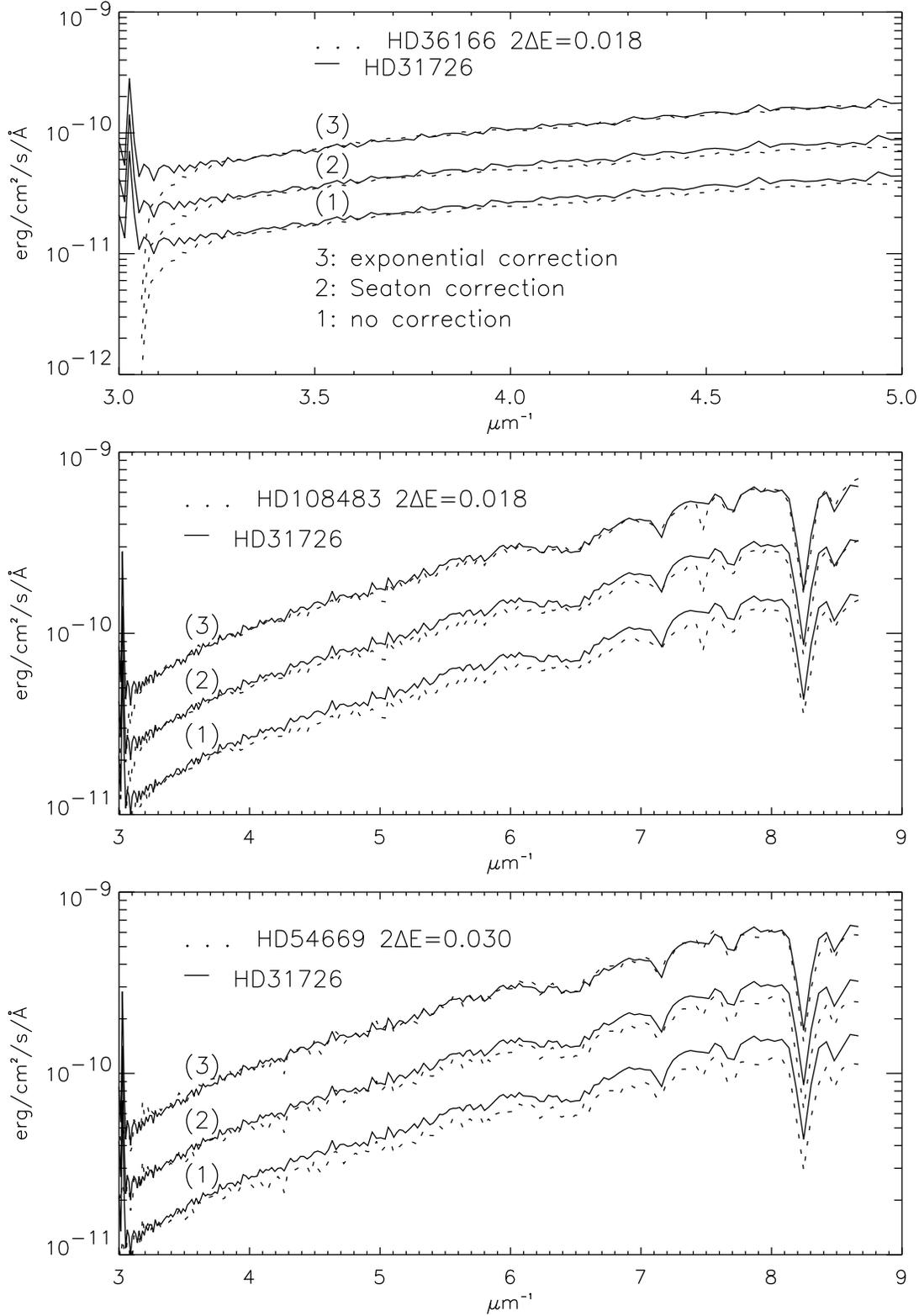}} 
\caption{Each plot compares the UV spectrum of HD31726 (plain line)
to the UV spectrum of another star of the sample (in dots). The 
bottom lines (1) compare the spectrum of HD31726 and the spectrum of 
the star scaled to the same value at small $1/\lambda$.
In the middle lines (2) the spectrum of the star is corrected for the 
difference of reddening with HD31726 with a Seaton's extinction law 
and the reddening difference found in the optical 
(section~\ref{analyvis}).
For the top lines (3) the correction factor is $\propto e^{-2\Delta 
E/\lambda}$.
$\Delta E$ is adjusted to give the best fit of the star to the 
spectrum of HD31726.} 
\label{fig:uv1}
\end{figure*}

\begin{figure*}[p]
\resizebox{!}{1.5\columnwidth}{\includegraphics{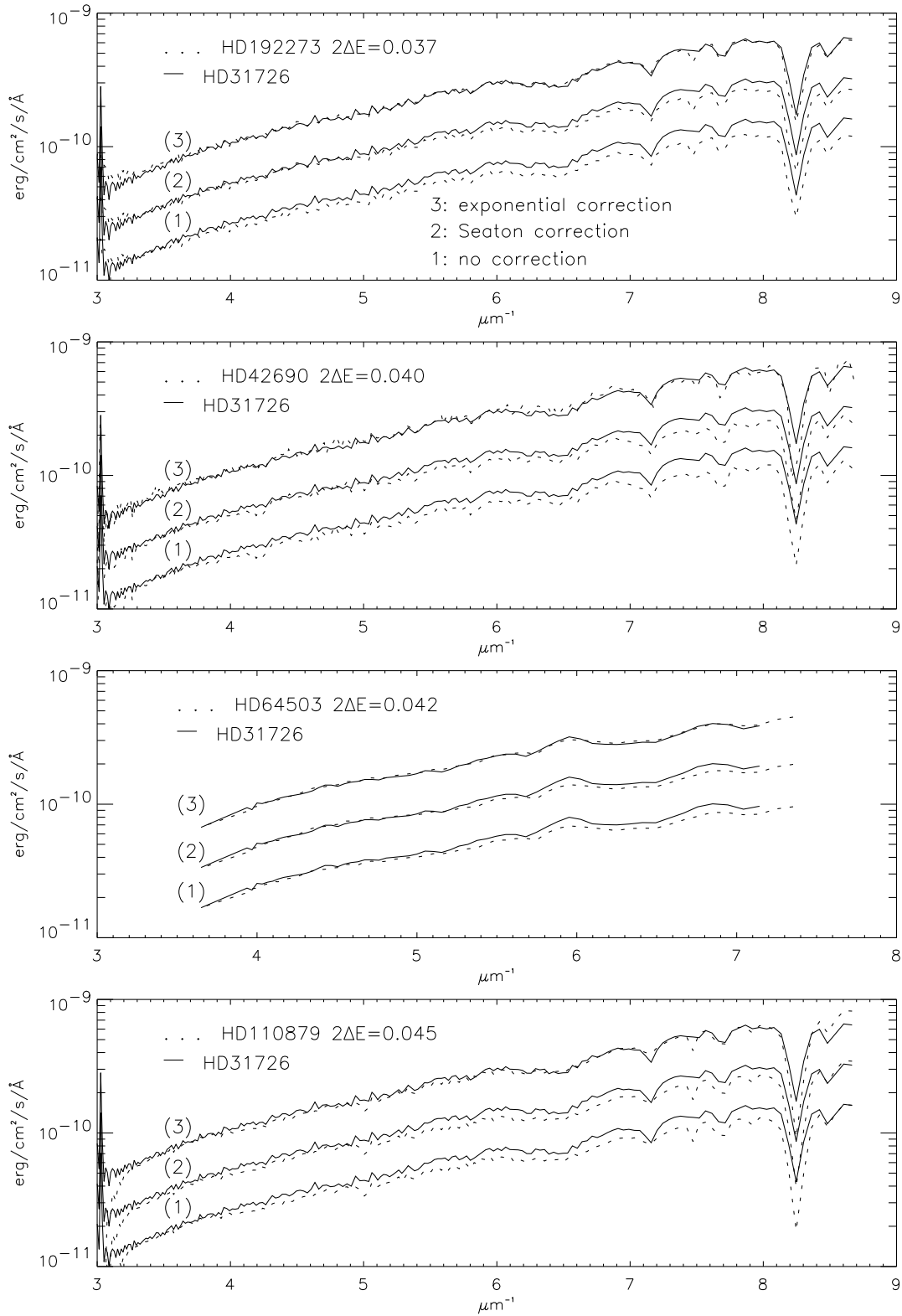}} 
\caption{See figure~\ref{fig:uv1}. For HD42690 
$E(B-V)$ used for curve (2) is $2\Delta E/2=0.02$ (the optical gives 
$E(B-V)=0$).} 
\label{fig:uv2}
\end{figure*}

\begin{figure*}[p]
\resizebox{!}{1.5\columnwidth}{\includegraphics{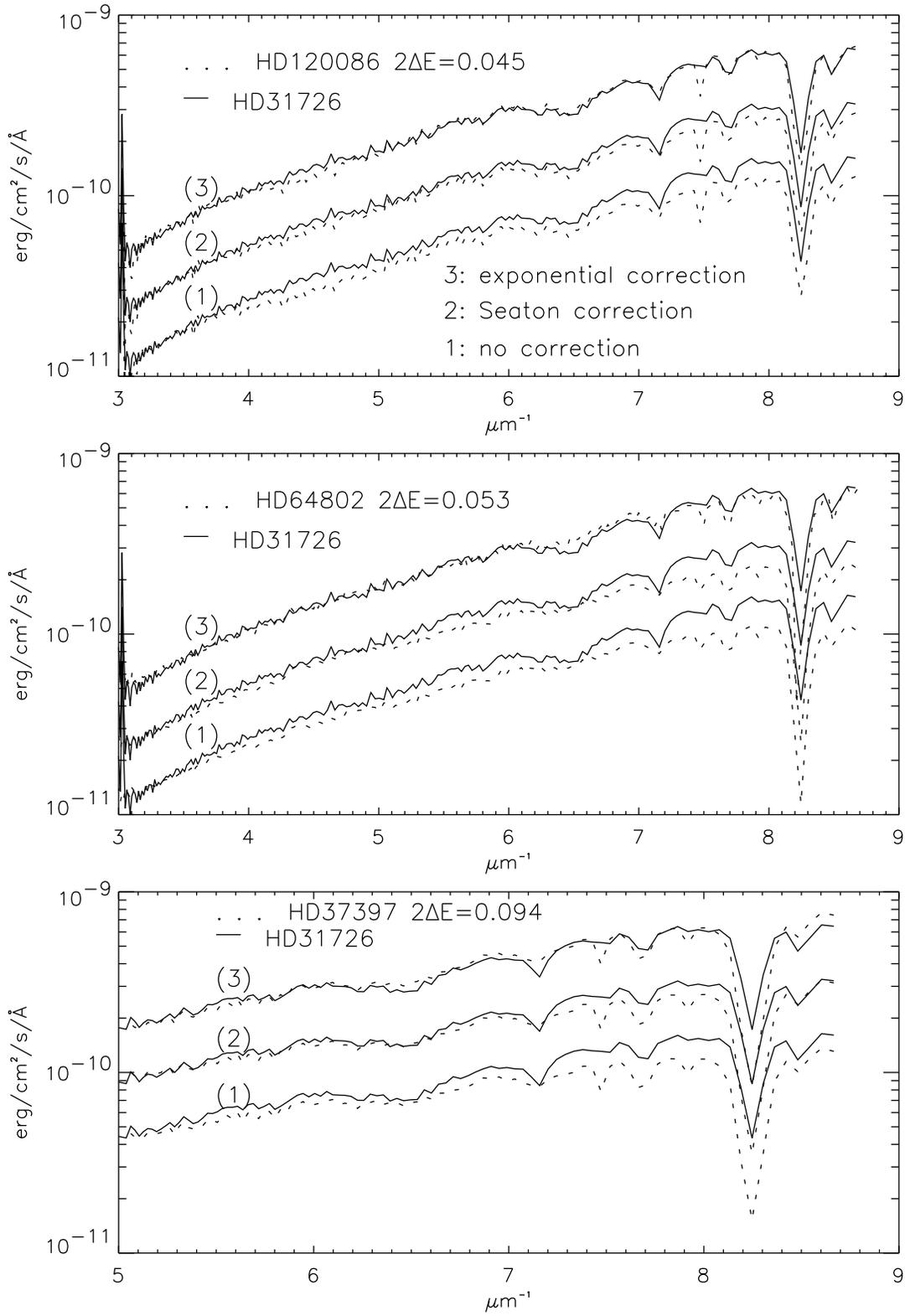}} 
\caption{See figure~\ref{fig:uv1}. } 
\label{fig:uv3}
\end{figure*}

Curves (1) in figures~\ref{fig:uv1} to \ref{fig:uv3} are, for each star, the spectrum of the 
star and the spectrum of HD31726, scaled to the same value at small 
$1/\lambda$.
The y-axis of the figure is in log-coordinate.

For all stars, the spectrum of the star progressively spreads away from 
the spectrum of HD31726.
The correction to apply to the spectrum of each of the star to fit 
the spectrum of HD31726 is an exponential of $1/\lambda$.

The curves marked (3) on the plots are the spectrum of 
HD31726 (plain line) and, for each star, the spectrum of the star
multiplied by an exponential $e^{-2\Delta E/\lambda}$, 
with $2\Delta E$ an appropriate exponent.
The two spectra superimpose well for all the stars.
The choice of $2\Delta E$ was guided by the value $2\Delta E_{vis}$ 
which is found in the visible (next section) and is more constrained.
The correction was then optimized.
For each star the best value found for $2 \Delta E$ is reported in 
table~\ref{tbl:stars}, column $2\Delta E_{UV}$.
The uncertainty of the values is large, but e.g. less than $\sim \pm 0.01$, except for the two stars with half a 
UV spectrum (HD36166 and HD37397), for which the values are given with a tolerance of $\pm 
0.02$.

For the stars of lowest reddening, HD122980 and HD36166, the 
UV slope can not be estimated accurately.
HD122980, with a spectrum very similar to the spectrum of HD31726, is 
not represented.
The difference of slope between the spectrum of HD36166 and of 
HD31726 is small and the spectrum of HD36166 is limited in 
wavelength extent.
The optical value (section~\ref{analyvis}) found for these stars,
which fits in the possible range of values of the UV slope difference, was finally adopted.

\begin{figure}
\resizebox{!}{0.8\columnwidth}{\includegraphics{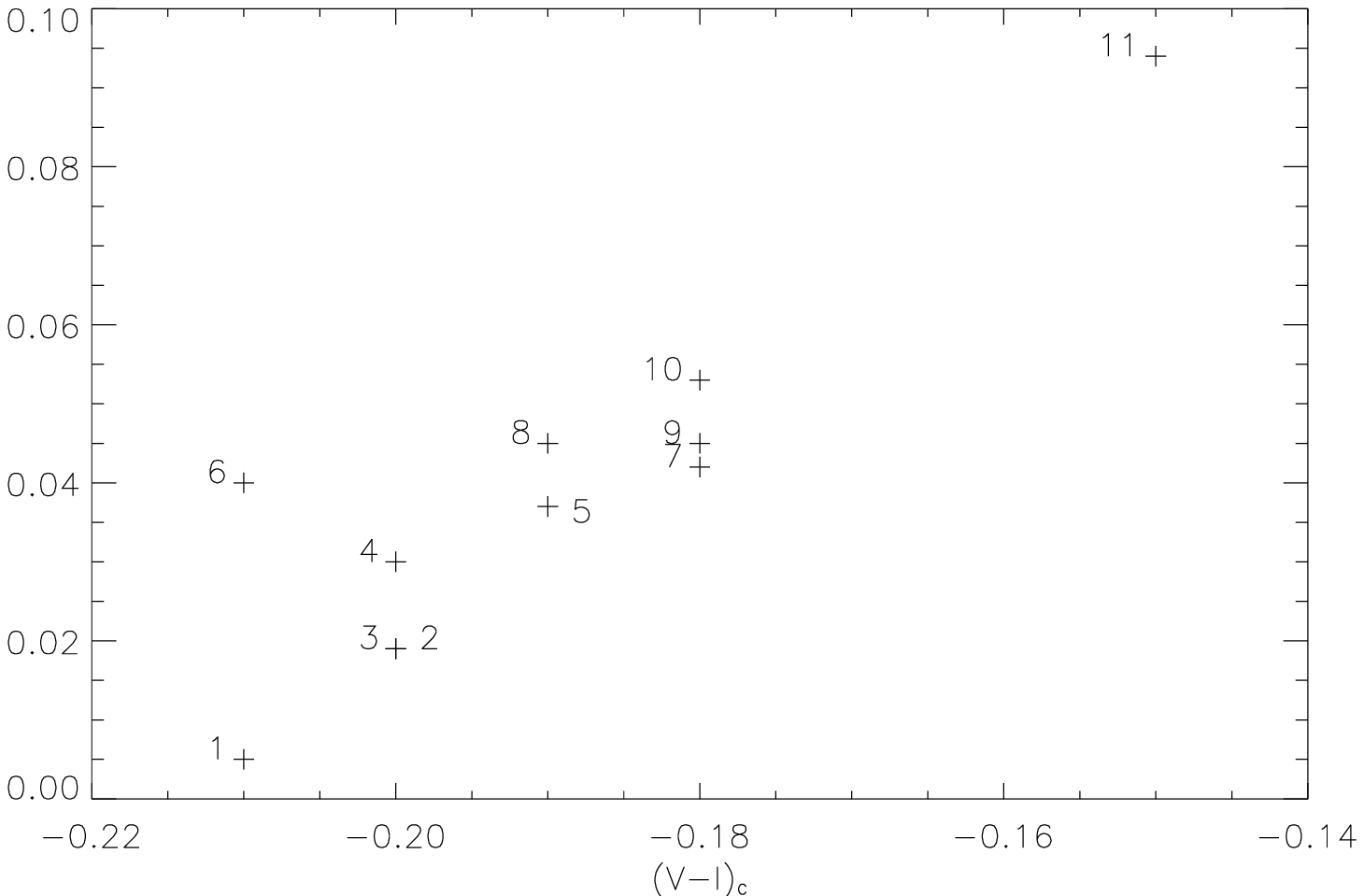}} 
\caption{For each star the exponent $2\Delta E_{uv}$ (from 
table~\ref{tbl:stars}) is 
plotted against V-I (table~\ref{tbl:stars}).
} 
\label{fig:vmi_uv}
\end{figure}

Figure~\ref{fig:vmi_uv} plots the slope $2\Delta E$ found for each star as a function of 
$V-I$.
It is clear that the increase of the UV slope is related to the 
increase of $V-I$, hence to the increase of $E(B-V)$.
\subsection{Optical data} \label{analyvis}
In the optical the magnitudes were not converted back to fluxes.
For each star the $V$, $R$, $I$ magnitudes were 
compared to the magnitudes of HD31726 on a magnitude-wavelength diagram.
The $B$ magnitudes are set to 0, which is equivalent to a scaling of 
the spectra at the same value in the $B$-band.
This comparison shows that the line drawn by 
the magnitudes of the star bends down more and more with increasing 
$E(B-V)$. 
The correction to apply to the magnitudes of the star 
so that they best match the magnitudes of HD31726 is a linear function of $1/\lambda$:
$1.08\times 2\Delta E_{vis}(1/\lambda -1/\lambda_B)$.

\begin{figure*}[]
\resizebox{!}{1.5\columnwidth}{\includegraphics{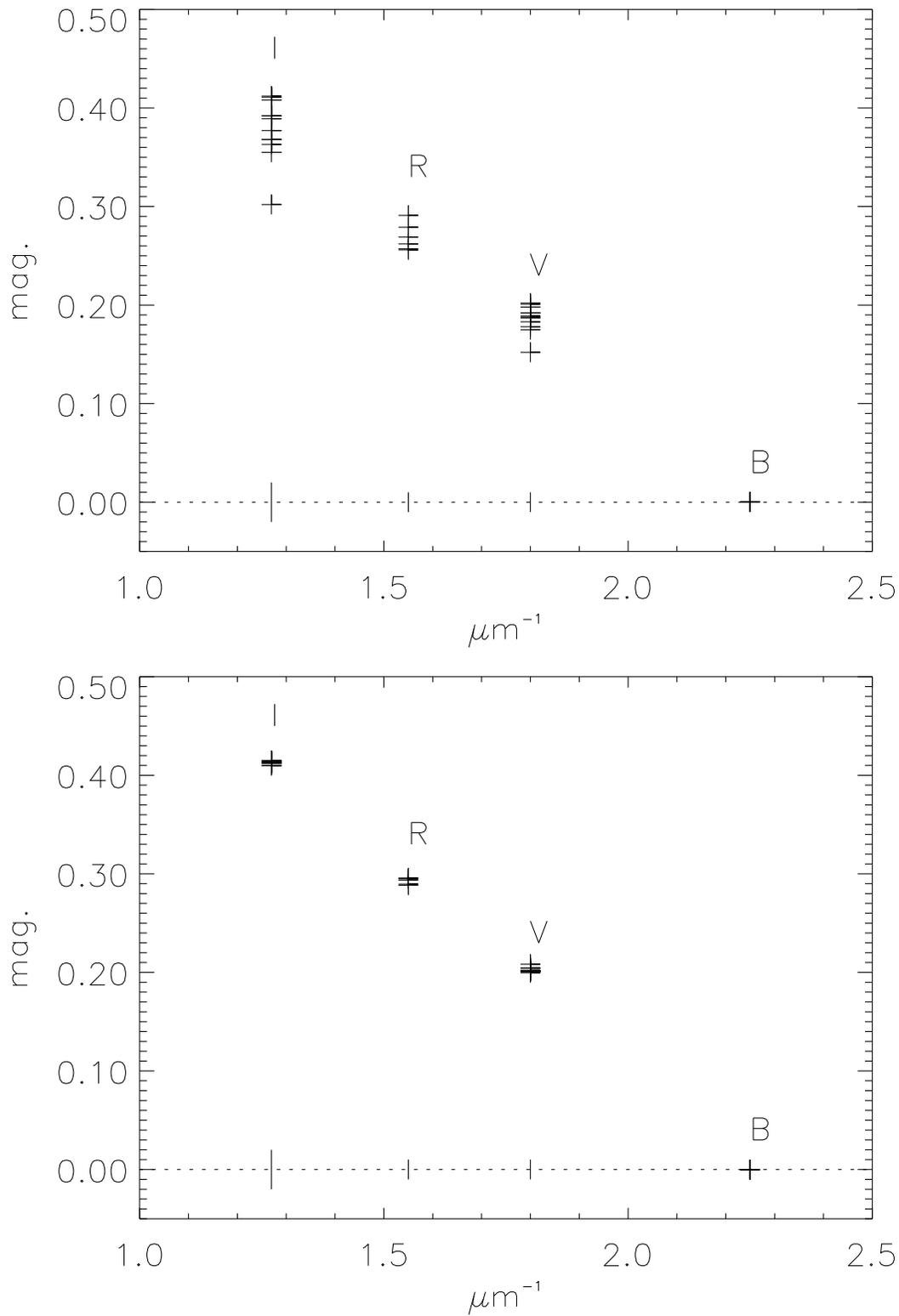}} 
\caption{\emph{top}: the magnitudes of the stars are plotted on the 
same figure. 
They are all scaled to 0~mag. in the $B$-band.
The mean error on the magnitudes is shown by the vertical bars.
\emph{bottom}: the stars are corrected for their difference of 
reddening compared to HD32617.
The correction brings all magnitudes, in each photometric band, to equality within the error 
margin.} 
\label{fig:vis}
\end{figure*}

I chose not to reproduce the one to one comparison and instead to 
group all the magnitudes, before and after correction, on the same plots.
Figure~\ref{fig:vis}, top, plots the magnitudes of all the stars as a 
function of wavelength.
The spread of the magnitudes increases with increasing wavelength and cannot be explained by the error bars 
(the vertical lines on the plot estimate an error of $\pm 0.01$, $\pm 
0.01$, $\pm 0.02$ for the $V$, $R$, $I$ bands).

The magnitudes of the 
stars after correction are plotted on the bottom plot of figure~\ref{fig:vis}.
The agreement between the corrected magnitudes is very good.
The error on $2\Delta E_{vis}$ should be less than 0.005.
The value of $2\Delta E_{vis}$ used in the plot is reported in 
table~\ref{tbl:stars}.
\subsection{Comparison of $2\Delta E_{vis}$ and $2\Delta E_{uv}$} \label{specomp}
\begin{figure}
\resizebox{!}{0.8\columnwidth}{\includegraphics{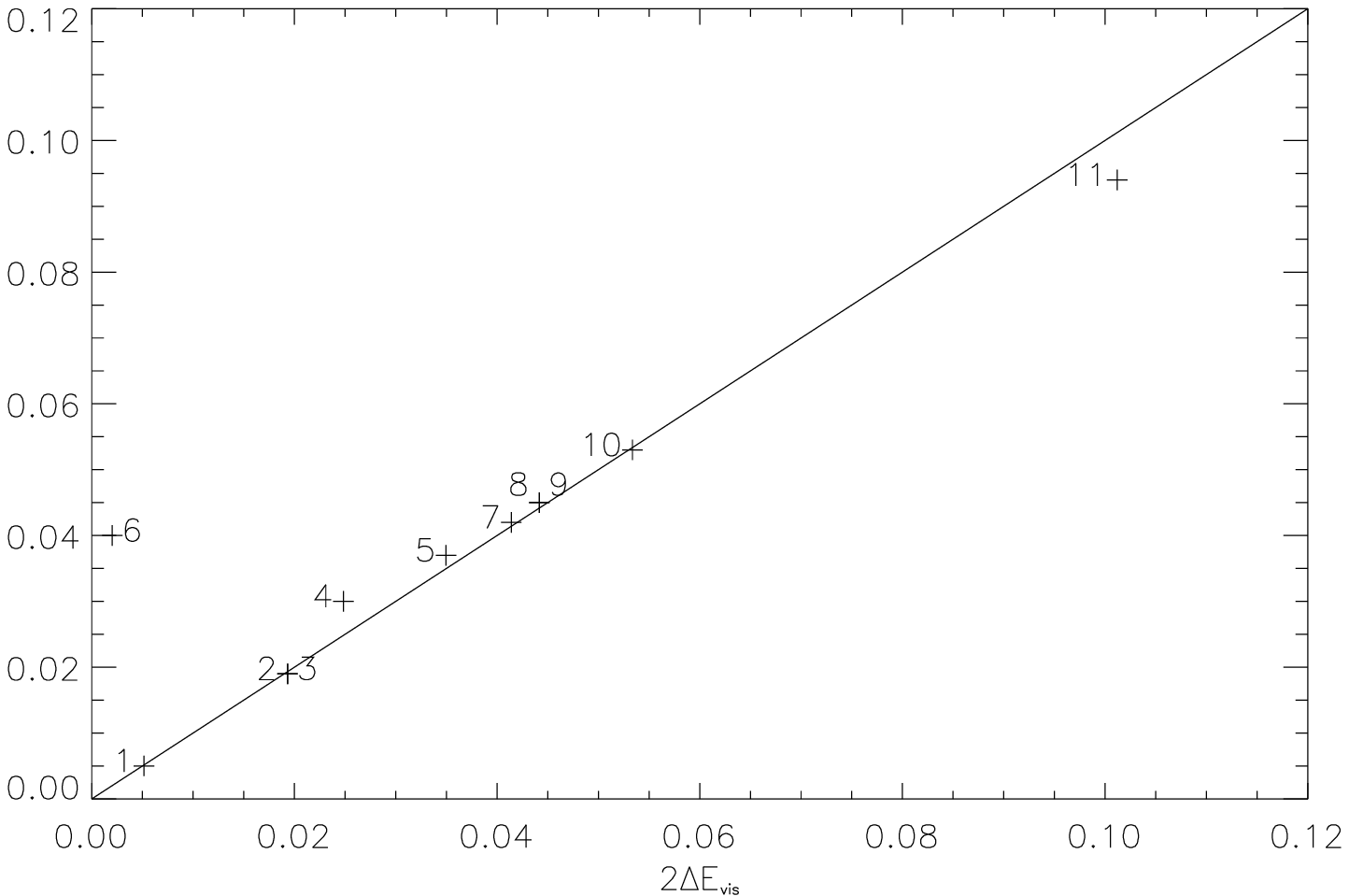}} 
\caption{The exponent found in the UV and in the visible are plotted 
one against the other.
Numbers label the stars (table~\ref{tbl:stars}).
} 
\label{fig:vis_uv}
\end{figure}

The UV and optical slopes, $2\Delta E_{uv}$ and $2\Delta E_{vis}$ 
found in the preceding sections are plotted 
one against the other in figure~\ref{fig:vis_uv}.
For all stars except one, HD42690, the agreement between the UV and optical data 
is good:
there is equality, within the error bars, 
between the exponents which were found in the optical and 
in the UV.

The $B-V$ and $I-V$ colors of HD42690 are close to these of HD31726 and
do not indicate an excess of reddening compared to HD31726,
while the IUE UV spectrum of the star is less steep than HD31726.
The IUE spectrum of HD42690 agrees with the TD1 spectrum so that 
-instead of accepting a particular extinction law for HD42690- I 
would rather question the optical data for the difference of 
exponent found in the UV and in the optical.
\subsection{Extinction curves} \label{extcur}
\begin{figure*}[p]
\resizebox{\hsize}{!}{\includegraphics{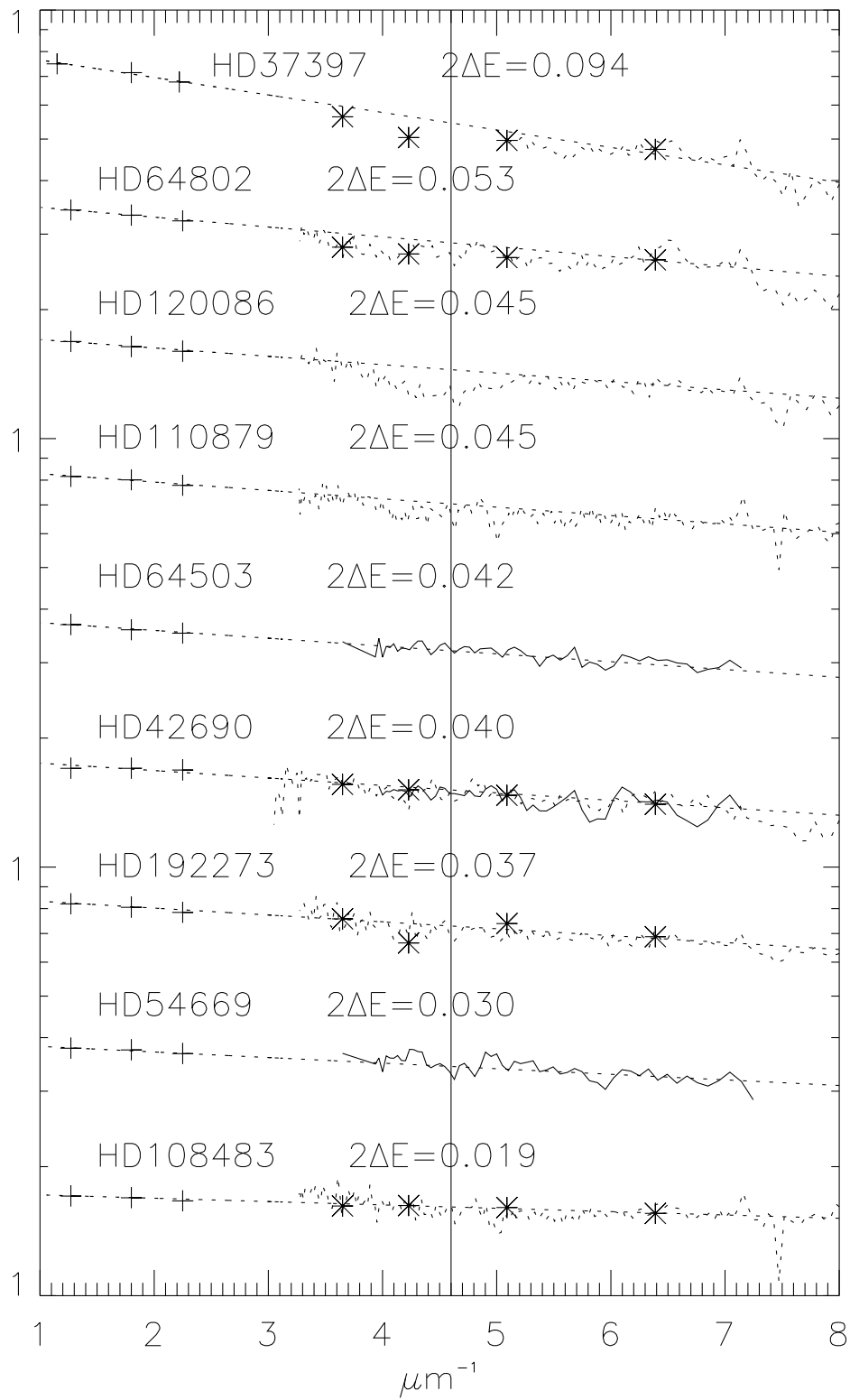}} 
\caption{The spectrum of each star is divided by the spectrum of 
HD31726. 
The resulting spectra are scaled by an arbitrary factor.
Dotted spectra correspond to IUE spectra, plain lines to TD1 spectra, `*' 
to TD1 photometry, `+' to visible photometry.
The straight lines are exponentials of $1/\lambda$ with the 
exponents written above the curves, after the name of the stars.
The vertical line shows the position of the bump at $4.6\,\mu\rm m^{-1}$.} 
\label{fig:extcur}
\end{figure*}
\begin{figure*}[p]
\resizebox{\hsize}{!}{\includegraphics{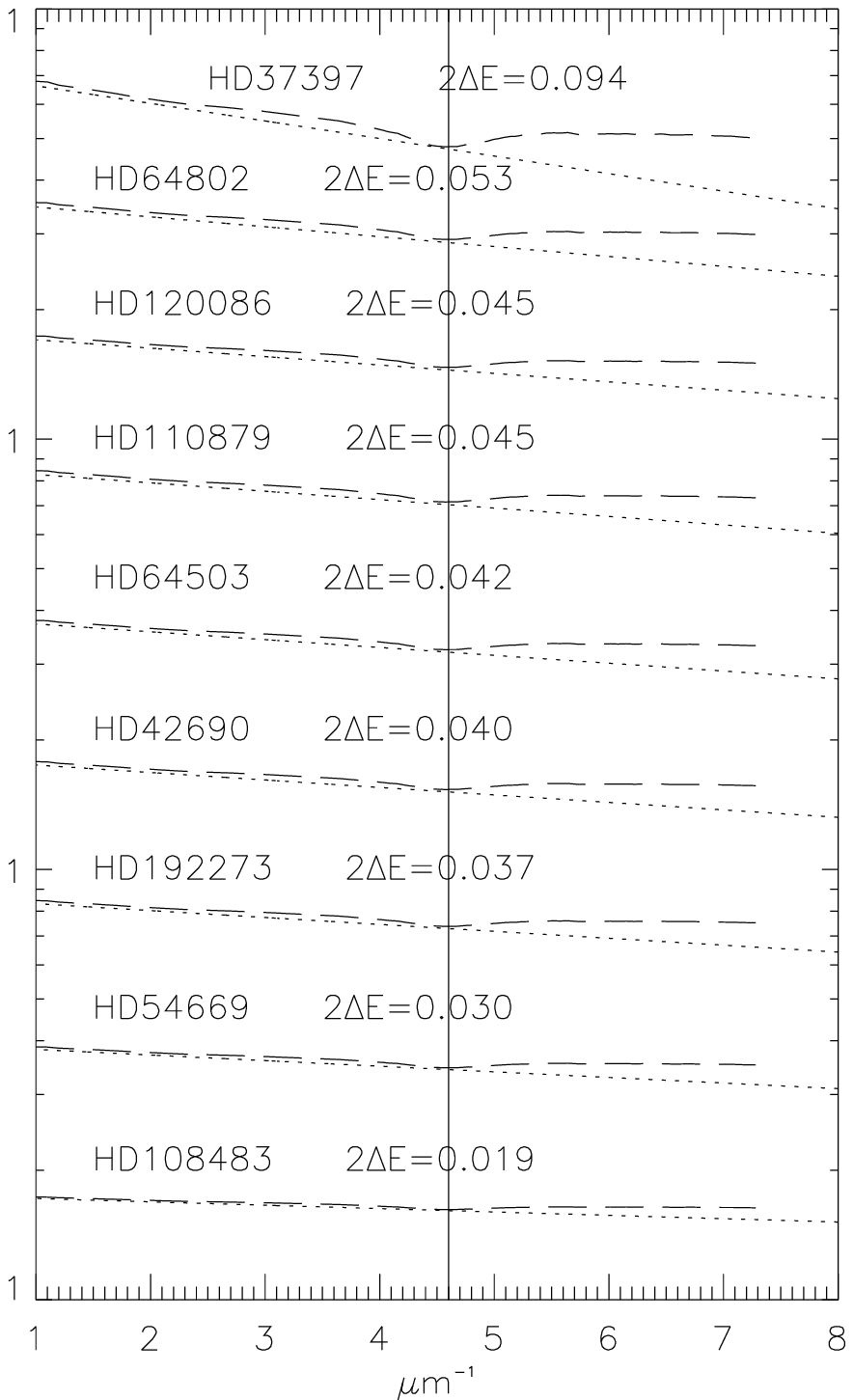}} 
\caption{This figure is similar to figure~\ref{fig:extcur}.
For each star only the slope, straight line, is kept from 
figure~\ref{fig:extcur}.
In dashes the curves predicted by Seaton's extinction 
law for $\Delta E(B-V)=\Delta E$.
The curves coincide in the optical, but the standard theory predicts 
a different and smaller slope in the UV.}
\label{fig:extcur1}
\end{figure*}
Figure~\ref{fig:extcur} plots the ratio (spectrum of a star)/(spectrum 
of HD31726) versus $1/\lambda$.
`*' on the curves correspond to the photometric TD1 fluxes 
which were added when necessary.
IUE spectra (in dots) were e.g. used except for HD54669 and HD64503 were 
TD1 spectra (plain lines) were 
preferred.
HD122980 and HD36166 were excluded from the plot.
The small $E(B-V)$ difference with HD31726, added to a limited 
wavelength range of the UV spectrum for HD36166, makes the curves too 
noisy to differenciate any slope for these stars.

For some of the curves a multiplicative factor was necessary to adjust the 
optical and the UV curves.
This factor is always close to 1, between 0.93 and 1.08.
It is within the uncertainties we can expect from the optical and UV calibrations.

The curves are scaled by an arbitrary factor to fit on the same plot.
The straight lines are the exponentials $e^{-2\Delta E_{uv}}$.
The name of the star and the exponent are written above each spectrum.
They are ordered by increasing $2\Delta E$.

Each curve approaches what was called a `reduced spectrum' of the 
star in preceding papers, analogous to the extinction curve in the 
direction of the star. 
The curves exactly represent reduced spectra 
if HD31727 is corrected for reddening, a correction 
which is discussed in section~\ref{refspec} but was not applied here.

For most of the curves there is no sign of a bump 
at $4.6\,\rm \mu m^{-1}$.
The spectrophotometric TD1 data of HD192273 may indicate the beginning 
of a bump.
It is also difficult to confirm the presence of a bump for HD110879.
Two stars, HD120086 and HD37397, with the highest slopes, do have a bump.
It is not clear whether HD64802, with a slope in between the ones of HD120086 and HD37397, has 
or not a bump.
\section{Discussion} \label{dis}
\subsection{Origin of the variations of slope of the spectra} 
\label{redori}
Beside extinction, variations of the intrinsic properties of each star, such as the temperature, 
can also produce a change in the slope of the spectrum.
The Kurucz's model, runned for $T=22000^{\circ}$~K and for 
$T=24000^{\circ}$~K (all other parameters are equal) give spectra which differ 
by $\alpha e^{0.025/\lambda}$ ($\alpha$ a constant).
For the spectrum to be more substantially modified the temperature needs to 
be changed by amounts which probably imply a change of spectral type.
An exponent of $0.07$ will be found for a temperature difference 
$\Delta T=4000^{\circ}$~K.
The spectral type of each star is well 
established, which rules out too large temperature differences 
between the stars.
A change of the temperature also modifies the spectral features, the Balmer 
jump, hence the 
factors ($\alpha$) which are necessary to adjust the spectra in the optical and 
in the UV. 
Hydrogen and Helium spectra will e.g. be modified which does not seem to be the 
case for the stars of the sample (figures~\ref{fig:uv1} 
to \ref{fig:uv3}).

The other explanation of the observed variations of slope of the spectra is reddening.
With IRAS images it is easy to verify that all the stars are behind 
interstellar clouds.
They are often on the edge of a cloud, which explains the low column 
densities in the directions of the stars.
The $100\,\mu$m IRAS emission at the stars' locations, compared 
to a minimum at a position close to the star, is often found to be $\sim 
1\,$~MJy/sr (it is larger when there is a local heating), confirming
low column densities.

The intrinsic color of a B2V star is $-0.24$~mag \citep{johnson}.
If a reddening of $E(B-V)\sim 0.05$ is adopted for HD31726, the reddening of 
the stars of the sample stands between $0.05$ and $0.1$.
$A_V$ will range from $0.15$ to $0.3$.
$I_{100}/A_V$, the ratio of IRAS $100\,\mu$m surface brightness to 
$A_V$ found in the litterature ranges from $18.6$~MJy/sr/mag \citep{bou} at the pole 
to much lower values towards the galactic plane ($5.5$~MJy/sr/mag in 
the direction of the Polaris Flare, \citet{zagury99}).
In the directions of the stars selected for this paper the expected infrared 
surface brightness at $100\,\mu$m will be between a few tenth and a 
few MJy/sr.

Although it is not possible, from IRAS emission,
to fix precisely the dust column density in the 
directions of the stars, the observed $100\,\mu$m 
emission agrees with the values found for $E(B-V)$, assuming 
the slope variations are due to reddening.  

Lastly, the emergence of the bump feature when the slope $2\Delta E$ of the 
star increases (section~\ref{extcur} and figure~\ref{fig:extcur})
also points to reddening as the explanation of the variation of slope from 
star to star.
\subsection{Extinction law in very small column density mediums} 
\label{extlaw}
A doubt will remain on whether the slope variation 
from spectrum to spectrum is due to extinction or to the intrinsic 
properties of the stars.
For the reasons exposed in the preceding section, 
I favour the reddening as an explanation of these differences.

If so, the observations analysed in section~\ref{analys}, indicate that the 
spectra of B2V stars with very little reddening can be deduced 
one from the other by multiplication by an appropriate exponential 
of $1/\lambda$ in the optical and in the UV.
It also can be deduced that the same exponent will apply to the UV and 
to the optical since within the error margin there is equality between the exponents $2\Delta E_{vis}$ 
and $2\Delta E_{uv}$.
If continuity exists between the two wavelengths range, 
the same function, $\alpha e^{-2\Delta E/\lambda}$ ($\alpha$ a 
constant), transforms one spectrum into the other, from the 
optical to the UV.
The constancy of the $\alpha$ coefficient was discussed in section~\ref{extcur} 
and is ensured within the calibration problems which are expected 
when different sets of data are compared.
Small differences of $\alpha$ from the optical to the UV, due to variations from star to star of the Balmer jump, 
are tolerable, but must remain small.

The implication is that the extinction law in directions 
of very small column densities is a straight line from the optical to the 
far-UV.
\subsection{Confrontation of observations with the standard theory of extinction} 
\label{sttheo}
According to the standard theory different types of interstellar 
grains are responsible for the extinction of starlight in the 
optical on one hand and in the UV on the other.
Each type of grain has a different extinction law in the UV and in the 
optical, and the extinction in a given wavelength range is radically 
different from type to type (\citet{desert} and 
\citet{greenberg}).
It is then difficult to explain the continuity of the 
extinction, from the optical to the far-UV, which we arrived at in 
section~\ref{extlaw}.

The average extinction curve given by the standard theory 
(Seaton's extinction curve) can be used to predict the extinction 
in the UV if the optical extinction is known.
According to the standard theory, there is an important diminution of 
extinction in the UV compared to the extension of the optical 
extinction (figure~\ref{fig:seaton}).
Curves (2) on figures~\ref{fig:uv1} to \ref{fig:uv3} are the UV spectrum of HD31726 (plain 
line) and 
the spectrum of the stars corrected for reddening by Seaton's law. 
The slope of the corrected curves is still significantly 
lower than the spectrum of HD31726.

Another view of the problem encountered by the standard theory is given 
in figure~\ref{fig:extcur1}.
For sake of clarity the data points have been excluded from the plot, 
but we know from figure~\ref{fig:extcur} that they follow the slopes 
given by the dotted lines.
The dashed lines are the curves predicted by Seaton's extinction law for
an adopted reddening $E(B-V)=\Delta E$.
The dotted and dashed lines agree in the optical, while the slope 
predicted by the standard theory is much lower and clearly different in the UV, and does not 
correspond to the observations.
\subsection{Alternative explanation} 
\label{exptheo}
The only alternative explanation to the standard theory
was originally proposed by \citet{savage75}  and discarded on the basis of 
the arguments I have discussed in UV2.
It implies that an appreciable amount of starlight is scattered at close angular 
distance from the star and is added to the direct starlight.
The light received from a star is to be separated into two additive 
components, direct and scattered light.

The spectrum of stars with little reddenning ($E(B-V)$ a few 0.1)
naturally splits into these two components (UV2).
The scattered light appears in the far-UV where extinction, and the 
number of photons available for scattering, is higher.
Scattered light is superimposed on the tail of the exponential 
decrease of the direct starlight.
For these stars, the exponential decrease of the direct starlight is observed down 
to the bump region, farther than predicted by the standard theory.

If extinction is extremely low ($E(B-V)$ a few 0.01), as for the stars selected for this 
paper, scattered light is repelled even further to larger optical 
depths and becomes a negligible part of the UV light we receive from the direction of the 
star.
The observed extinction approaches the extinction of the direct starlight. 
According to section~\ref{extcur} it is linear in $1/\lambda$ in the optical and in the UV.
This is the law most commonly predicted by the Mie theory for large 
grains or by the 
theory of extinction by small grains.
 
An additional test between the two explanations of the extinction 
curve consists in comparing the ratios of the flux of the 
stars to the flux of HD31726 in two separate wavelengths domains.
Small uncertainties on the flux or magnitudes of the stars will not 
affect the comparison of ratios $F_\star /F_{ref}$, with $F_\star$ 
the flux of the star and $F_{ref}$ the flux of HD31726, if  the 
wavelengths where these ratios are compared are moved away as far as 
possible.
The $I$ band and the far-UV represent the two extrems of the 
wavelength range considered in this paper and available from existing 
data set.

In the $I$ band, the ratio $F_\star /F_{ref}$ is $r_I=10^{-(\Delta V 
-\Delta (V-I))/2.5}$ with $\Delta V $ and $\Delta (V-I)$ the 
differences in $V$ and $V-I$ magnitudes between the star and HD31726.
In the far UV the ratio $r_{uv}$ was found for each star by adjusting the UV 
spectrum of the star to the spectrum of HD31726 around 
$1/\lambda=7\,\mu\rm m^{-1}$.

From figure~\ref{fig:seaton} the difference of 
extinction 
$A_\lambda-A_I$ between the $I$-band ($1/\lambda=1.27$) and $7\,\rm \mu m^{-1}$ 
expected by the standard theory is $\sim 6.2E(B-V)$.
The standard theory predicts a ratio $r_{uv}/r_{I}\sim 10^{-6.2\Delta 
E/2.5}$ (dotted line on the plot).

If linear extinction continues in the UV, we expect:
\begin{equation}
    A_\lambda-A_I= 
    \frac{1/\lambda-1/\lambda_I}{1/\lambda_B-1/\lambda_V} E(B-V)
    =2.12(\frac{1\mu\mathrm{m}^{-1}}{\lambda}-1.27) E(B-V)
    \label{eq:alambda}
    \end{equation}
With $1/\lambda=7\rm\,\mu m^{-1}$, we deduce: 
$r_{uv}/r_{I}\sim 10^{-12.2\Delta E/2.5}$.

\begin{figure*}
\resizebox{!}{0.8\columnwidth}{\includegraphics{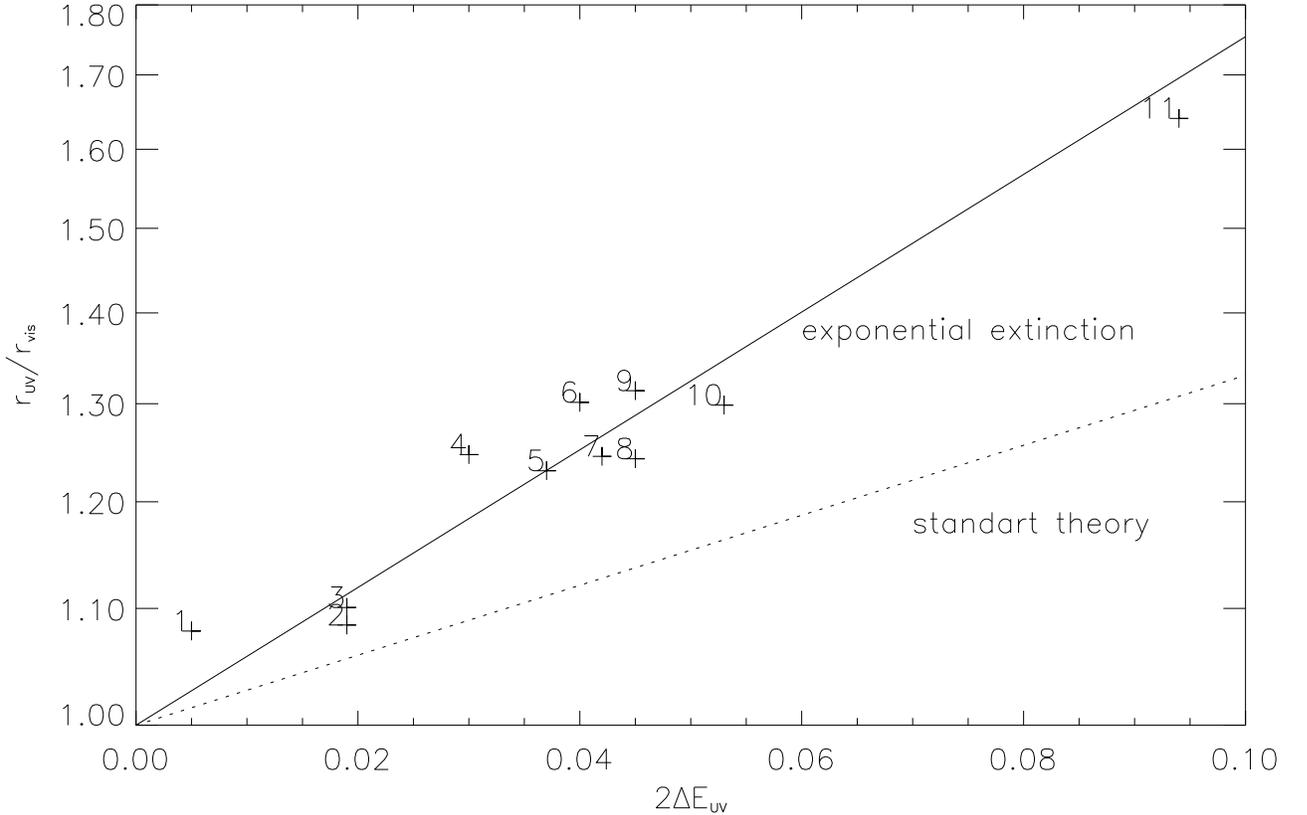}} 
\caption{The x-axis is the UV exponent found in 
section~\ref{analysuv} and reported in table~\ref{tbl:stars}.
The y-axis is the ratio $r_{uv}/r_{vis}$ defined in section~\ref{exptheo}
The plain line is $r_{uv}/r_{vis}$ predicted by a linear extinction 
over all the spectrum.
The dotted line is the prediction made from Seaton's extinction curve.
} 
\label{fig:uvsuri}
\end{figure*}

Figure~\ref{fig:uvsuri} plots $r_{uv}/r_{I}$ against $2\Delta E_{uv}$.
The observations closely follow the line corresponding to the linear 
extinction, for all stars.
\section{The spectrum of a non reddened star} \label{refspec}
From section~\ref{extlaw} we deduce that the spectrum of a B2V star 
corrected for extinction is proportional to the spectrum of HD31726 
multiplied by $e^{2E(B-V)_0}$, with $E(B-V)_0$ the reddening of HD31726.
$E(B-V)_0$ is close to $0.04$ with an uncertainty less than $\pm 0.01$.

The extinction law found in section~\ref{extlaw} cannot depend 
on the spectral type of the stars.
It follows that the unreddened spectrum of a star 
of a given spectral type is obtained from the spectrum of a star of 
same spectral type and of low reddening (a few 0.01 at most) by 
multiplication by $e^{2E(B-V)}$, with $E(B-V)$ the reddening of the 
star.

The adquirement of the spectrum of a non reddened star from the 
spectrum of a slightly reddened one was already used in UV3 to obtain 
a reference spectrum for the star HD46223.
It is justified here.
\section{Conclusion} \label{conc}
In this paper I have studied the extinction of stars of same spectral 
type in a few directions of space with very little reddening.
The spectra of these stars can be deduced one from the other
by an exponential $e^{-2\Delta E /\lambda}$.
This dependence is verified and continuous from the near-infrared to 
the far-UV.

The difference between the spectrum of the stars can be due either to 
the intrinsic properties of the stars or to extinction.
Differences of temperature from one star to another may provoke a 
small variation of the slope of the star but will also modify the 
H and He spectra of the star.
Too large temperature differences will also imply a change of 
spectral type.
On the other hand all the stars are behind interstellar clouds of low 
column densities and the $E(B-V)$ value of the stars agree with the 
estimated $100\,\mu$m surface brightness of the 
clouds.
When the difference of the slope of the star with the slope of HD31726 
becomes large enough a bump appears at $4.6\,\mu\rm m^{-1}$, which is 
an additional sign that the difference of slope between the spectra is due to 
reddening.

If it is assumed that the differences of slope of the spectra 
presented here are due to extinction, 
the extinction law in directions of low reddening is 
a straight line from the near-infrared to the far-UV.

This extinction law is difficult to reconcile with the standard 
theory of extinction.
According to the standard theory interstellar extinction is due to 
the combined effect of three types of grains, each of which dominates 
the extinction in a specific wavelength range (\citet{greenberg} for a 
recent review).
The optical extinction due to large grains reaches a ceiling in the 
near-UV, while the far-UV extinction is attributed to very small 
grains.
Within this framework it is not possible to understand how very 
low reddenings are explained by a single extinction law which extends 
from the near-infrared to the far-UV.

The main result of the paper is that the standard theory of 
extinction is contradicted by the observation of stars with low 
reddening.

The extinction of the stars in regions of low 
reddening coincides with the idea that the spectrum of a star 
can be contaminated by starlight scattered at close angular distance 
from the star.
If the reddening is important enough ($E(B-V) >\sim 0.1$), substantial 
scattered light is received in the far-UV (UV2).
Further increase will see the scattered light appear in the near-UV 
and in the optical.
If the reddening is very low, the scattered light is negligible: 
the light we receive is mainly the direct starlight extinguished by 
the extinction law of interstellar dust.
The linear extinction law which is found here is in agreement with the 
basic theory of extinction by dust particles.

The present attempt to clarify the relationship which exists 
between the spectrum of stars of same spectal type and low reddening 
leads to the adquirement of the spectrum of a non reddened star of given 
spectral type, from the spectrum of a star with very small reddening.
The star must be only slightly reddened so that its' spectrum is not 
yet contaminated by scattered light;
the reddening $E(B-V)$ of the star must be less than a few 0.01.
The correction to apply to the spectrum of the star is 
$e^{2E(B-V)/\lambda}$.

{}
\end{document}